\documentclass[aps,prl,twocolumn,groupedaddress]{revtex4}

\usepackage{graphicx}
\usepackage{dcolumn}
\usepackage{bm}

\begin{document}

\title{ Spectrum of $\pi$-electrons in Graphene As a Macromolecule}

\author{Lyuba Malysheva}
\affiliation{Bogolyubov Institute for Theoretical Physics, 03680 Kyiv, Ukraine}
\author{Alexander Onipko}
\email[]{aleon@ifm.liu.se}
\affiliation{Bogolyubov Institute for Theoretical Physics, 03680 Kyiv, Ukraine}

\date{\today}

\begin{abstract}
We report the exact solution of the spectral problem for a graphene sheet framed by two armchair- and two zigzag-shaped boundaries. The solution is found for the $\pi$ electron Hamiltonian and gives, in particular, a closed analytic expression of edge-state energies in graphene. It is shown that the lower symmetry of graphene, in comparison with $C_{6h}$ of 2D graphite, has a profound effect on the graphene band structure. This and other obtained results have far-reaching implications for the understanding of graphene electronics. Some of them are briefly discussed. 
\end{abstract}


\maketitle


{\it Introduction}.---Most theoretical studies on the electronic properties of graphene start from either a H\"uckel-type Hamiltonian \cite{Nakada,Fujita,Saito,Peres1,prb2007}, or k/kp versions of the Dirac Hamiltonian \cite{Ando,Peres2,Brey,Silvestrov}; for a review, see \cite{Geim}. In one way or another, references are made to the symmetry points of 2D graphite band structure \cite{Wallace}. At these points the valence and conduction $\pi$ electron bands join each other, and the dispersion of electrons and holes is linear up to energies $<$ 1 eV. This and the periodicity of two sublattices of 2D graphite unites electrons and holes near the Fermi energy with massless fermions. However, for finite-size graphene structures such as graphene ribbons, the structure illustrated in Fig. 1 is more relevant as a reference model than the 2D graphite lattice. Shown here is a plane macromolecule consisting of $N\times\cal N$ hexagons which are arranged in sequences of $N$ ${\cal N}$-long oligomers of polyacene and coupled to each other via $\cal N$ C-C covalent bonds. All dangling bonds along the graphene edges are filled by hydrogen atoms. 

This model has been the focus of a number of works, but until now only approximate analytical solutions of the spectral problem have been proposed \cite{Nakada,Fujita,Brey}. To begin, we present the exact description of the graphene $\pi$ electronic structure. It is shown that the spectrum is fully determined by the dispersion relation (which is different from that known for 2D-graphite) supplemented by the generalized Lennard-Jones equation. This part of the description is substantially based on our previous studies of the band structure of conjugated oligomers \cite{jcp107,Chapter}. Next, we discuss applications of the obtained equations in the context of graphene and its daughter lattices, armchair and zigzag graphene ribbons (GRs) and carbon nanotubes (CNTs). In conclusion, we express the edge-state spectrum in terms of elementary functions and show its agreement with the exact results. In what follows, C-C hopping integral $t$ is the only parameter of the semi-empirical Hamiltonian of the graphene macromolecule. This parameter is used as a unit of energy.


{\it Exact solution of the eigenvalue problem for graphene}.---By taking an appropriate representation of the molecular orbitals of graphene, we reduce the initial two-dimensional Schr\"odinger problem to $\cal N$ independent sets of $2N$ one-dimensional equations. Each set describes a hypothetical oligomer consisting of $N$ monomers, as is illustrated in  Fig. 1 by dashed, numbered frames. 
\begin{figure}
\includegraphics[width=0.35\textwidth]{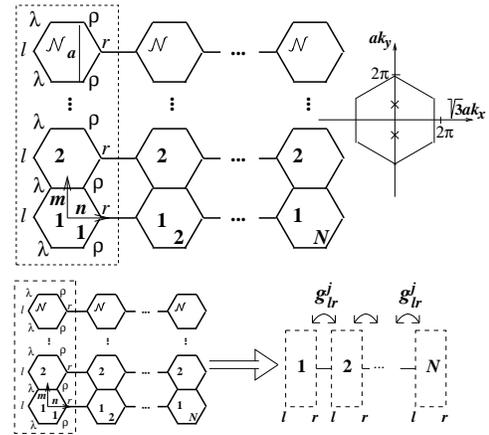}
\caption{Indication of labels of carbon atoms used in the present description of the $\pi$ electron spectrum of $N\times\cal N$  honeycomb lattice. Right inset shows symmetry points of 2D graphite (hexagon vertices) and graphene (crosses). In lower part, dashed-framed block (oligomer of polyacene with the length $a{\cal N}$) can be thought as a monomer M in an M-oligomer. The monomer Green's function matrix elements completely determine electron spectra of M-oligomers \cite{jcp107,Chapter}. In Eq.~(\ref{3}), the role of these matrix elements is played by $g_{l,r}^j$ and $g_{l,l}^j$.}
\end{figure}

Exploiting $m,n$, and $\alpha=l,\lambda,\rho,r$ labeling explained in Fig. 1, the $\pi$ electron wave function that satisfies the Schr\"odinger equation $H\Psi=E\Psi$ with the tight-binding Hamiltonian of $N\times\cal N$ graphene sheet, can be represented as follows
\begin{equation}\label{1}
\Psi = \sum_{m}\sum_{n}\sum_{\alpha=l,r,\lambda,\rho}
\psi_{m,n,\alpha} |m,n,\alpha\rangle,
\end{equation}
where $|m,n,\alpha\rangle$ is the $2p_z$ orbital at the $\alpha$th atom of benzene ring with coordinates $\{m,n\}$ with a summation running over all sites of the honeycomb lattice ($|m,n,\alpha\rangle=0$ if these lattice sites are empty, for example, $|1,N+1,\alpha\rangle=0$), and
\begin{equation}\label{2}
\psi_{m,n,\alpha}  =\sum_{j=1}^{\cal N} \phi_{n,\alpha}^j a_m^j, \quad \alpha = l,r.
\end{equation}
In the latter expansion, $\displaystyle a_m^j = \left(\sqrt{2}/\sqrt{{\cal N}+1}\right)\sin\xi_jm$, $\xi_j =\pi j/({\cal N}+1)$, $j=1$, 2, \ldots, ${\cal N}$, and coefficients $\phi_{n,\alpha}^j$, $\alpha = l,r$ are subjected to equations
\begin{equation}\label{3}
\phi_{n,\alpha}^j  =g_{\alpha,l}^j\phi_{n-1,r}^j+ g_{\alpha,r}^j \phi_{(n+1),l}^j,
\end{equation}
where $g_{l,r}^j =g_{r,l}^j$, $g_{l,l}^j=g_{r,r}^j $, ${\cal D}_j g_{l,r}^j =4\cos^2(\xi_j/2)$,
${\cal D}_j g_{l,l}^j =E[E^2-1-4\cos^2(\xi_j/2)]$, and zeros of ${\cal D}_j= [E^2-4\cos^2(\xi_j/2)]^2-E^2$ determine the spectrum of an $\cal N$-long acene.

The set of equations (\ref{3}) is central in this otherwise standard derivation. As already mentioned, it appears in the theory of M-oligomers, M-M-\ldots-M, where the energy dependent quantities of $g_{l,l}^j$ and $g_{r,l}^j$ are the monomer Green's function matrix elements referring to the same (left or right) or different binding atoms of a monomer M; see Fig. 1. This analogy was first noticed by Klymenko \cite{C1}. 

Finding the eigenvalues of (\ref{3}) and thus solving the eigenvalue problem det$(H-EI)=0$ yields 
\begin{equation}\label{5}
\cos\kappa = f(g_{l,l}^j,g_{l,r}^j)= \frac{1}{2g_{l,r}^j}\left[1+ \left(g_{l,r}^j\right)^2
-\left(g_{l,l}^j\right)^2\right],
\end{equation}
where $\kappa$ and $E$ are interrelated via
\begin{equation}\label{7}
\frac{\sin \kappa N}{\sin \kappa(N+1)}=
g_{l,r}^j \left [\left(g_{l,l}^j\right)^2 - \left(g_{l,r}^j\right)^2 \right ]^{-1}.
\end{equation}
Formally the same equations as (\ref{5}) and (\ref{7}) appear in the tight-binding description of M-oligomers \cite{jcp107,Chapter}. As a particular case, in the Lennard-Jones theory of polyenes (M = C=C) \cite{LJ}, the right hand side of Eq.~(\ref{7}) is independent of energy and equal to a certain constant. 

By taking into account the explicit expressions of $g_{l,l}^j$ and $g_{r,l}^j$, Eqs.~(\ref{5}) and (\ref{7}) can be transformed into
\begin{equation}\label{8}
E^{\pm\,2}=1\pm4\left|\cos({\xi_j}/{2})\cos (\kappa/2)\right|+4\cos^2({\xi_j}/{2}),
\end{equation}
and
\begin{equation}\label{9}
\frac{\sin \kappa^\pm N}{\sin \kappa^\pm (N+1/2)}=\mp2\cos(\xi_j/2),
\end{equation}
respectively. Within the interval $0\le\kappa^\pm\le\pi$, Eq. (\ref{9}) with sign plus or minus has $N$ solutions 
which determine $j$-dependent quantum numbers $\kappa^\pm_{j,\nu}$, $\nu =0,1,\ldots,N-1$ and hence, the spectrum of the graphene sheet, $E=\pm E^\pm_{\kappa^\pm_{j,\nu},j}$. To be precise, Eqs.~({\ref{8}}) and ({\ref{9}}) determine $4N\cal N$ of the total number $2N(2{\cal N}+1)$ of the $\pi$ electron levels. Additionally, there are two $N$-fold degenerate levels with energies $\pm1$. These can be proven to be the states with zero wave-function amplitudes at the $l$ and $r$ sites, thus making them of no interest here. Below, only $4N\cal N$ $\pi$ electron states, $|j,\nu\rangle$, are considered.


{\it Spectra of graphene daughter lattices}.---Equation (\ref{8}) remains valid for periodic boundary conditions (PBCs), in which case the $\pi$ electron spectrum is fully determined by this single equation, 
where $\kappa = \kappa_l = 2\pi l/N$, $l=0,$1,2,\ldots,$N$$-$1
and $\xi_j =  2\pi j/{\cal N}$, $j=0,$1,2,\ldots,${\cal N}$$-$1. Thus defined, the dispersion relation reads
\begin{eqnarray}\label{10}
[E^\pm(k_x,k_y)]^2=&&1\pm4\left|\cos({ak_y}/{2})\cos (\sqrt{3}ak_x/2)\right|\nonumber\\
&&+4\cos^2({ak_y}/{2}),
\end{eqnarray}
where the minimal translation distance $a$ is indicated in Fig. 1; the correspondence between the continuos variables in the dispersion relation and discrete quantum numbers in Eq.~(\ref{8}) is as follows: $\sqrt{3}ak_x\leftrightarrow\kappa$, $ak_y\leftrightarrow\xi_j$. 

For the system in focus, the usage of Eq.~(\ref{10}) should be restricted to the range $0\le \sqrt{3}|k_x|,|k_y|\le \pi/a$. It is easy to see that, within this range, there are only two points, $k_x=0,ak_y=2\pi/3$ and $k_x=0,ak_y=-2\pi/3$, where $E^\pm(k_x,k_y)=0$ (instead of six for 2D graphite \cite{Wallace}). Not far away from these points, the dispersion relation (\ref{10}) can be approximated by the familiar formula
\begin{equation}\label{10a}
E^\pm(k_x,k_y)=\pm\frac{\sqrt{3}a}{2}\sqrt{k_x^2+(|k_y|-2\pi/3a)^2},
\end{equation}
that is by a linear form of dispersion, if it is expressed in terms of the deviation of {\bf k} from the points of zero energy. The reduced number of these zero points is a direct consequence of the lower symmetry in graphene in comparison with 2D graphite.  

\begin{figure*}
\includegraphics[width=0.3\textwidth]{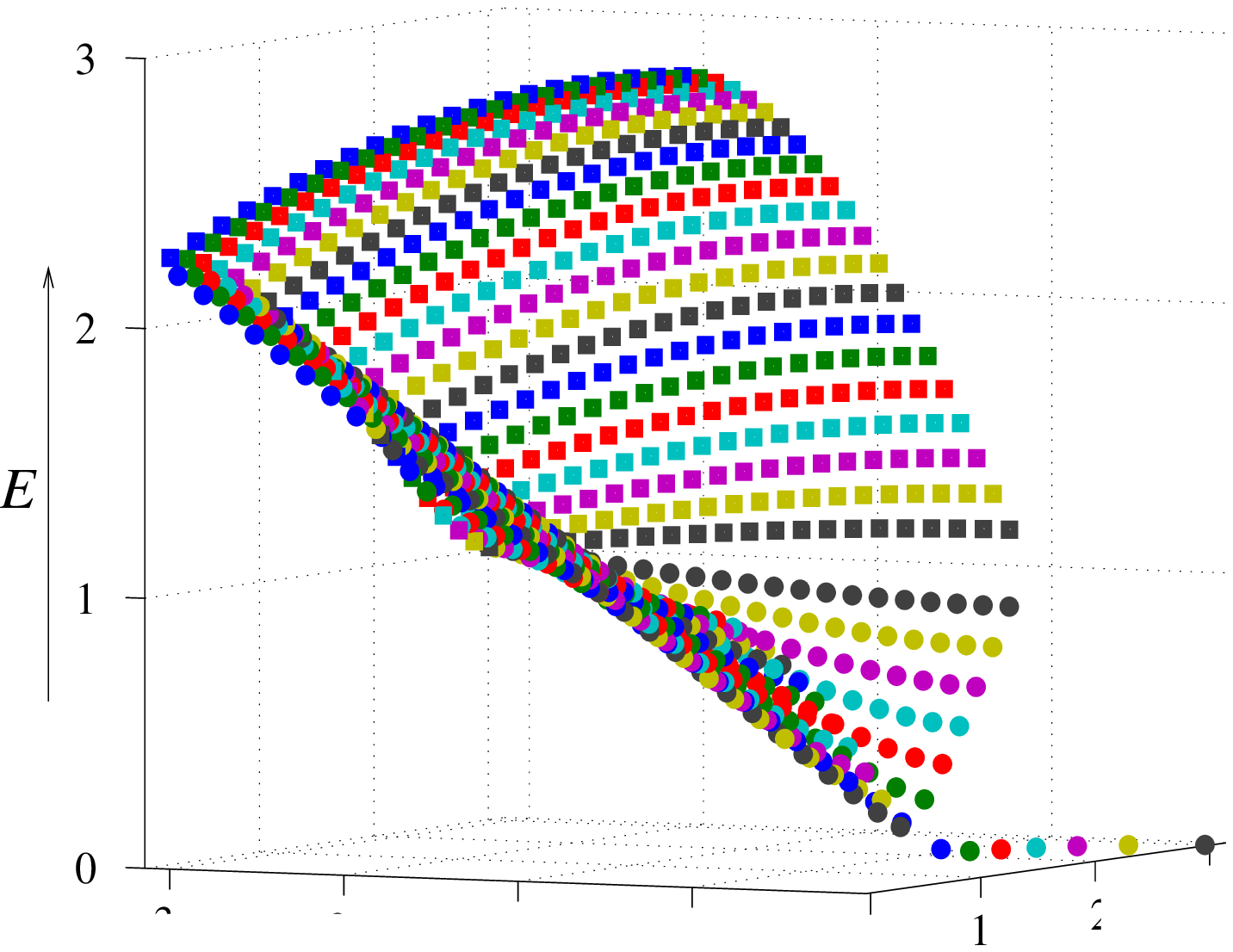}
\includegraphics[width=0.3\textwidth]{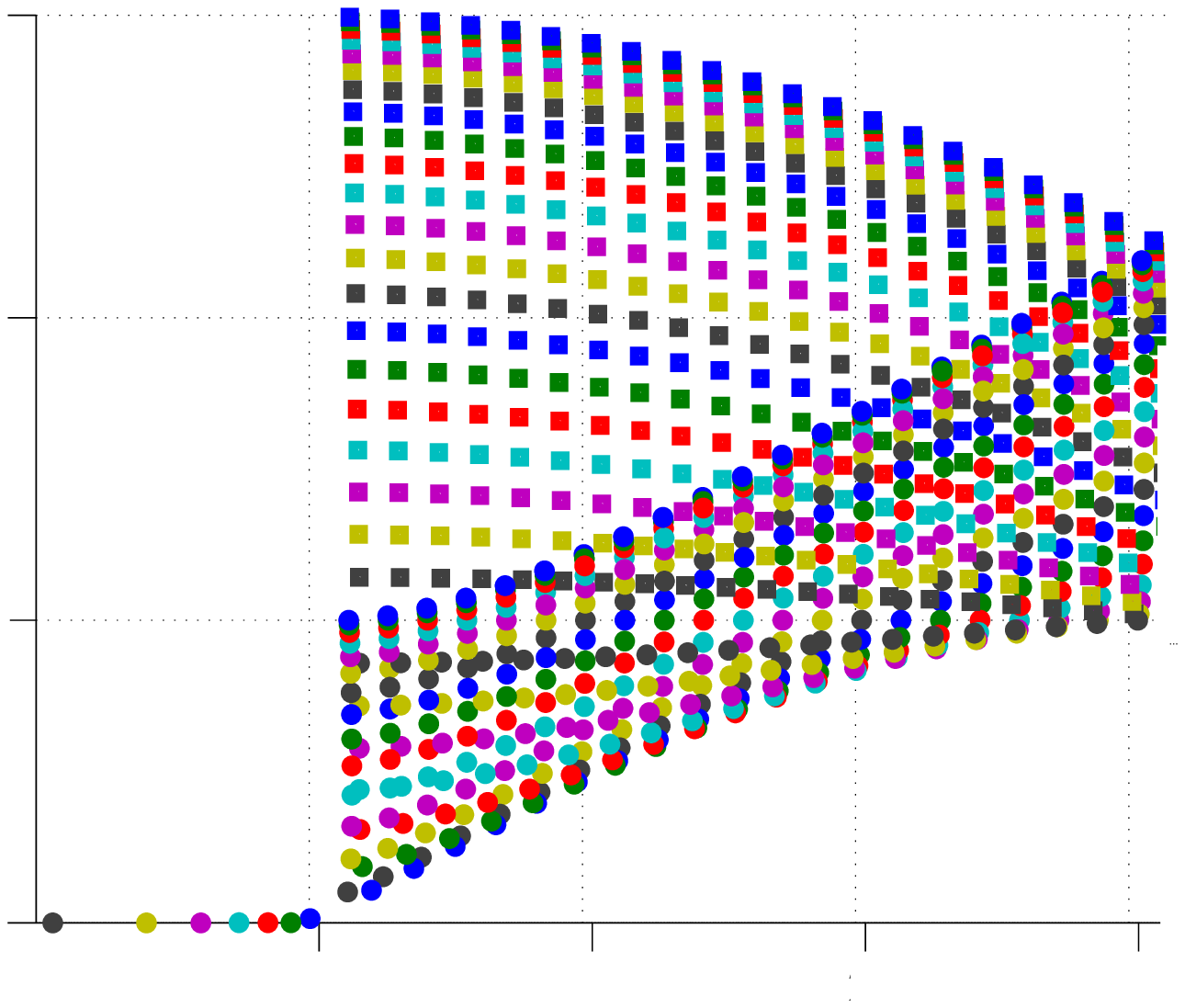}
\includegraphics[width=0.3\textwidth]{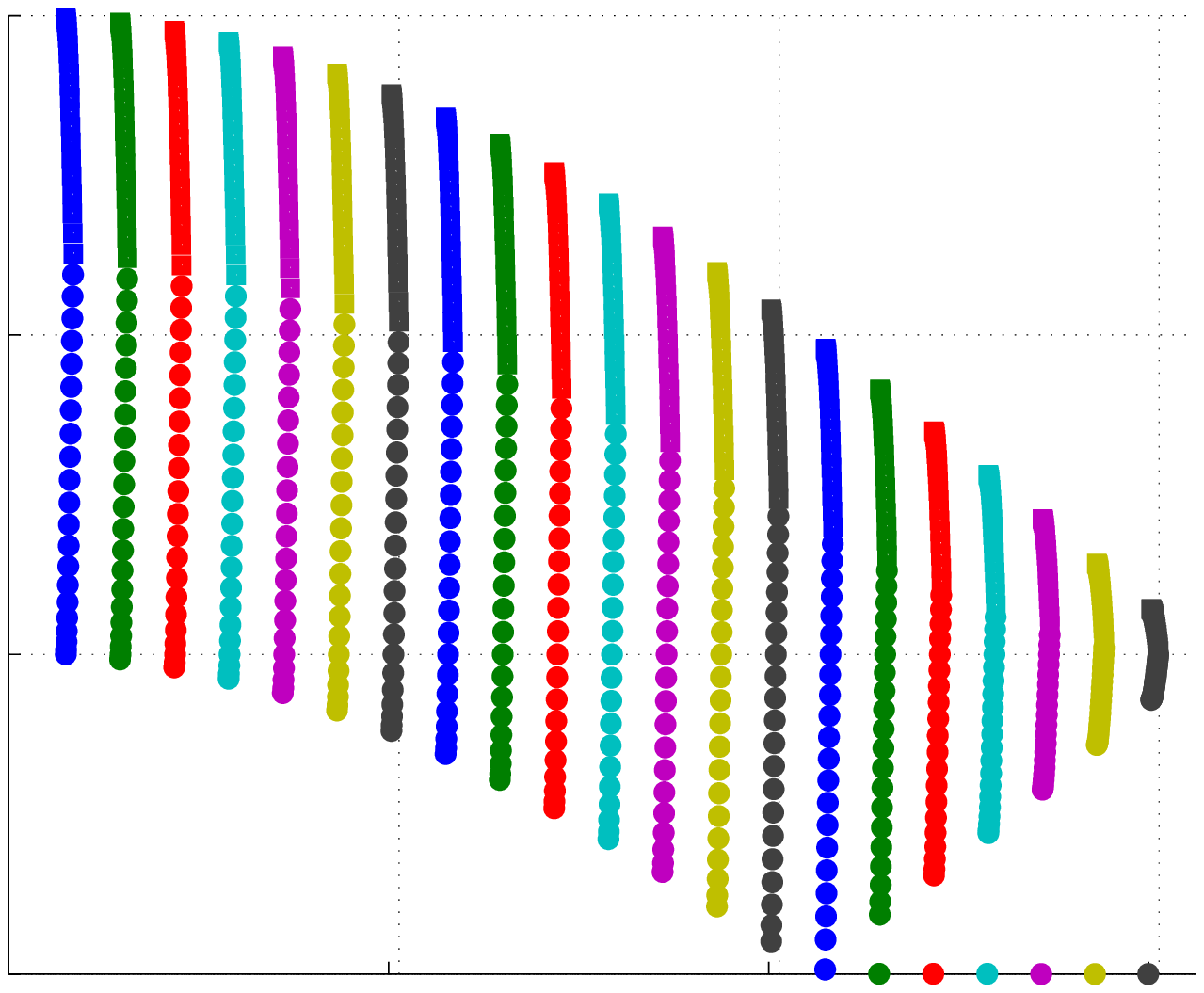}
\includegraphics[width=0.3\textwidth]{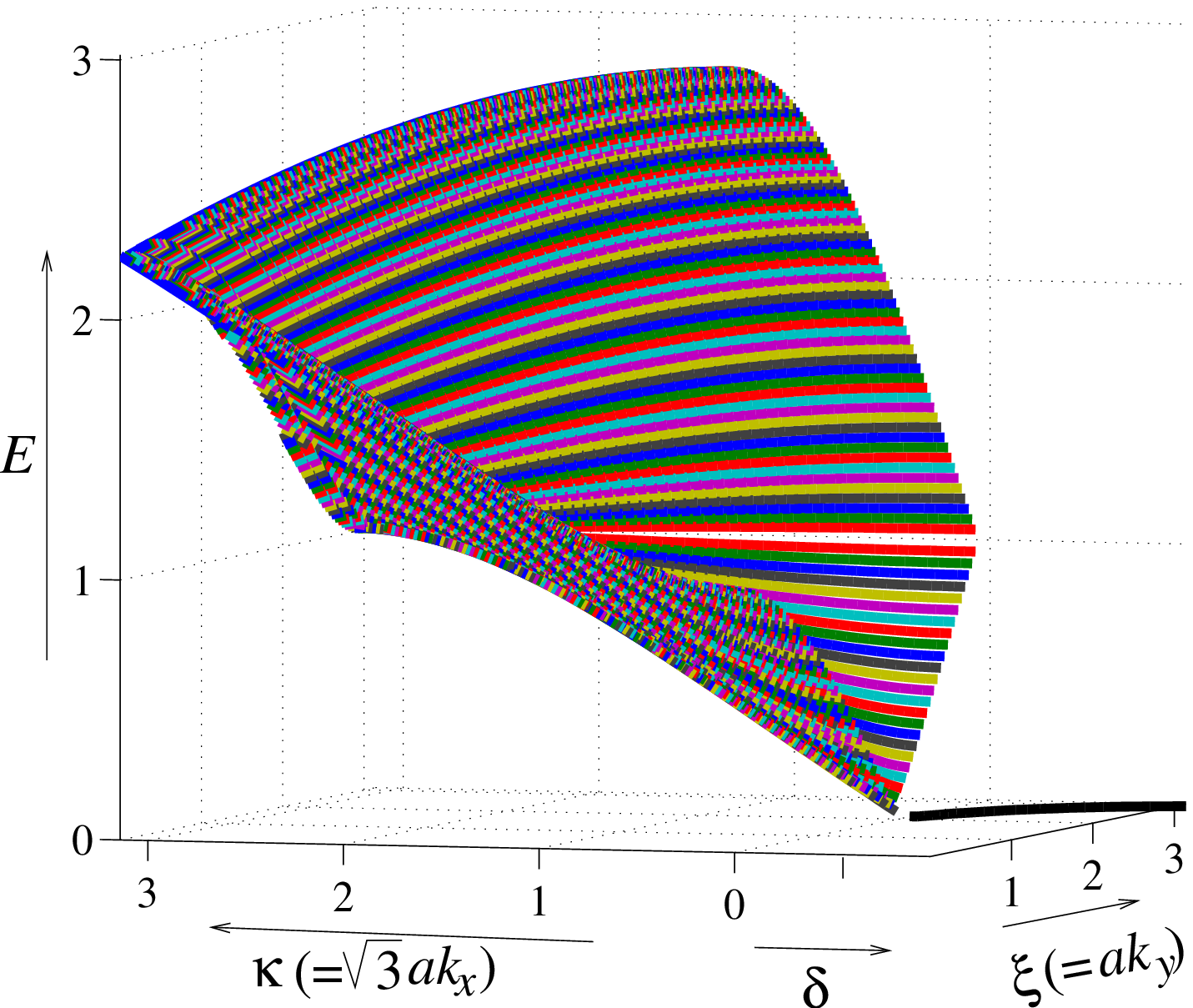}
\includegraphics[width=0.3\textwidth]{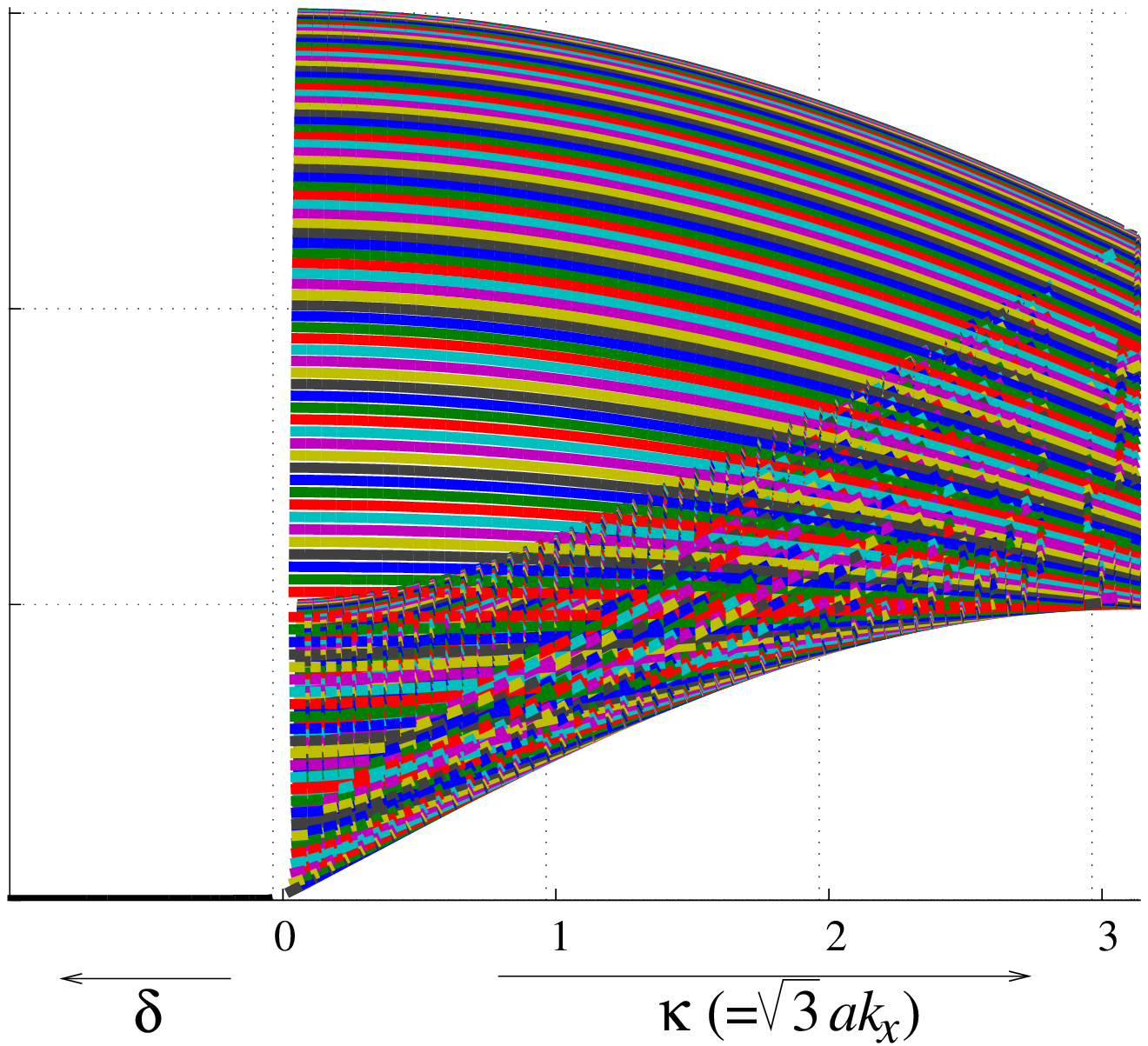}
\includegraphics[width=0.3\textwidth]{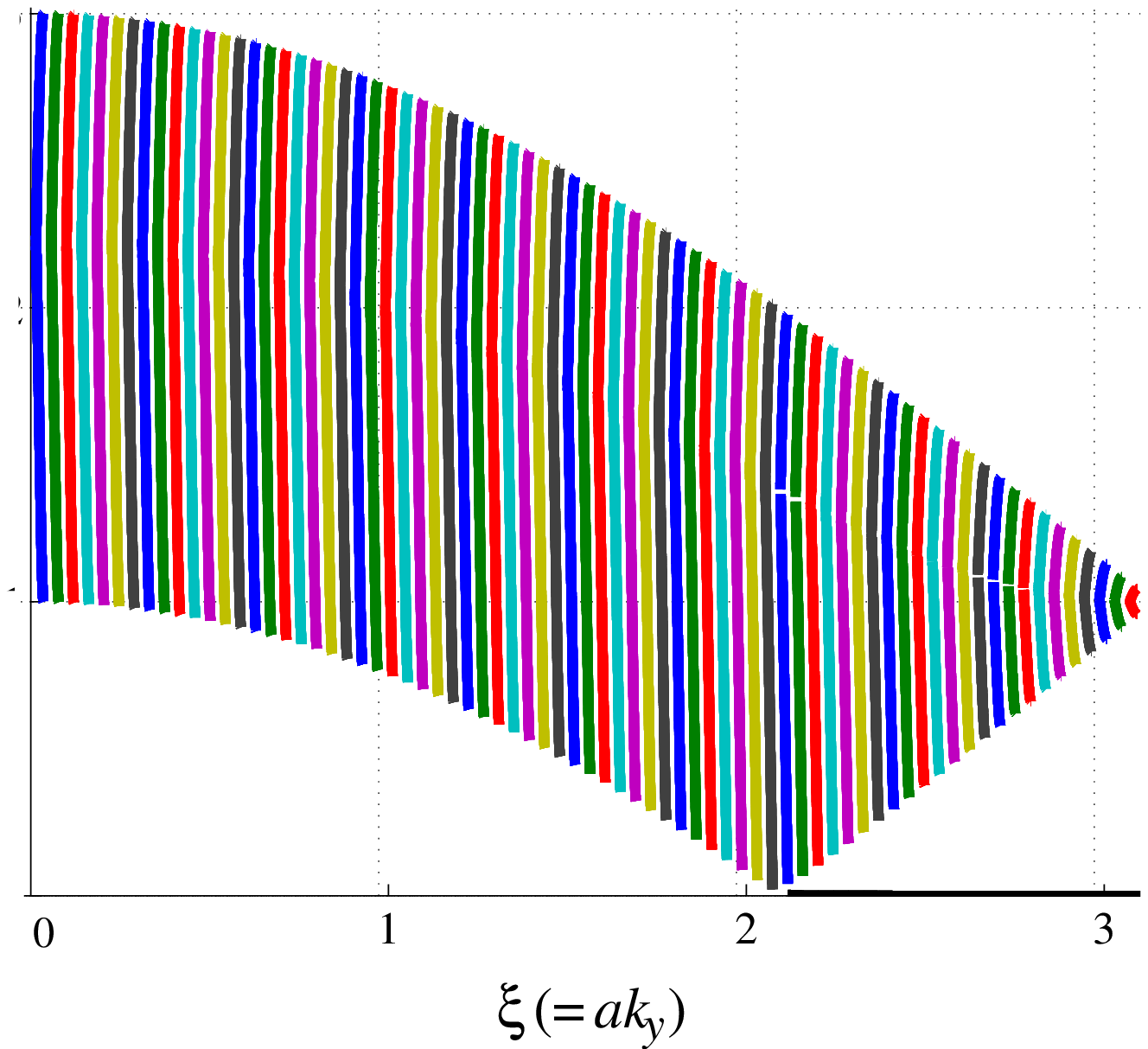}
\caption{Graphene spectrum $E^\pm_{\kappa^\pm_{j,\nu},j}$, $E\ge0$, according Eqs.~(\ref{8}) and (\ref{9}). Squares and circles correspond to signs plus and minus in these equations, as explained in the text; different colors indicate different values of $j$. From left to right: General view, projections $\xi=0$, and $\kappa = 0$. ${\cal N}=N=21$ and ${\cal N}=75$, $N=60$ for upper and lower panels, respectively. Imaginary quantum values of $\kappa$ are shown on $\delta$ continuation of $\kappa$ axis. The corresponding energy levels (edge-state levels) are seen as dots (upper panels) and black line (lower panels) in $\delta$-$\xi$ plane.} 
\end{figure*}

Equations (\ref{8}) and (\ref{9}) can be easily used for an instructive description of graphene daughter lattices, specifically, armchair and zigzag GRs and CNTs. However, the cases of finite and infinite (i.e., independent of the boundary conditions) systems must be clearly distinguished. For infinite armchair GRs, the spectrum is completely determined by Eq.~(\ref{8}), where $-\pi\le \kappa=\sqrt{3}ak_x\le \pi$. For infinite zigzag GRs, it is $\xi=ak_y$ that should be treated as a continuous variable in the two equations, (\ref{8}) and (\ref{9}). 

Formal description of {\it armchair} and {\it zigzag} CNT spectra is exactly the same, as that of {\it zigzag} and {\it armchair} GRs, respectively. The only difference is that the discrete quantum number is not determined by the open boundary conditions [as in Eq. \ref{2})] but by the PBCs. As a result, the use of the dispersion relation (\ref{10}) extends to the range $0\le \sqrt{3}|k_x|\le \pi/a$, $0\le |k_y|\le 2\pi/a$ and $0\le \sqrt{3}|k_x|\le 2\pi/a$, $0\le |k_y|\le \pi/a$, for zigzag and armchair CNTs, respectively. This means that the band structure of zigzag CNTs, as compared with armchair GRs, has two new points of zero energy, $E^\pm(k_x=0,k_y=\pm 4\pi/3a)=0$. In armchair CNTs, four special points appear, $(k_x=\pm2\pi/\sqrt{3}a,k_y=\pm 2\pi/3a)$. Further discussion of CNTs and GRs spectra can be found elsewhere \cite{Lyuba3}.

Some of the essential parts of the above discussion are exemplified in Fig. 2. Represented in its upper part is the $\pi$ electron spectrum of a 21$\times$21 graphene sheet, where each value of $j$ has its own color and circles and squares correspond to ``minus'' and ``plus'' branches, respectively, of Eq.~(\ref{8}). Three panels show (from left to right) energies of $j,\nu$ levels, and crossections of the spectrum by planes $\kappa$-$E$ and $\xi$-$E$. Levels which are represented by circles in $\delta$-$\xi$ plane correspond to imaginary values of $\kappa$. They associate with electron states which are localized near zigzag-shaped boundaries. The conjugated part of the spectrum, $j,\nu$ levels with negative energies, is the mirror reflection in $\kappa$-$\xi$ and $\delta$-$\xi$ planes. 

Transformation of the spectrum with the increase of $N$ and $\cal N$ is illustrated by the lower panels in Fig. 2. These give a visual representation of the graphene band structure $E^\pm(k_x,k_y)$. Note that the spectra shown in the mid and right lower panels have the same appearance as those which have been obtained in computational modeling of armchair and zigzag GRs \cite{Nakada}. 

{\it Spectrum of graphene edge states}.---By denoting that $\xi_j-2\pi/3=q_j$, it can be shown that Eq.~(\ref{9}) has $N$ real solutions,  $\kappa^-_{j,0}<\kappa^-_{j,1}$$<$\ldots$<$$\kappa^-_{j,N-2}<\kappa^-_{j,N-1}$, if $q_j<q^c$, where $q^c$ can be found from $E^-_{0,q^c}$ = $\pm [1-2\cos(\pi/3+q^c/2)]$. If $N>>1$ then, $q^c$ = $(\sqrt{3}N)^{-1}$. The smallest of the solutions for $\kappa$ becomes imaginary if $q_j>q^c$: $\kappa^-_{j,0}=i\delta_j$, $j\ge j^*$, where $j^*=[2({\cal N}+1)/3]$; $[A]$ denotes a minimal integer of rational number $A$. Note that $\kappa^-_{j^*,0}=\delta_{j^*}=0$, if $({\cal N}+1)/3$ is an integer. In Fig. 2, imaginary values of $\kappa^-_{j\geq j^*,0}=i\delta_j$ are shown on an extension of the $\kappa$ axis. Energy levels $E^-_{\kappa^-_{j\geq j^*,0}=i\delta_j,j}$ are very close but are never equal to zero; see below. 

The energies which satisfy Eqs.~(\ref{8}) and (\ref{9}), and fall into the interval $-(2N+1)^{-1}<E<(2N+1)^{-1}$, correspond to imaginary values of $\kappa^-_{j,0}$ and hence, to electron states, decaying towards the mid of the graphene sheet along the 
armchair direction. These states (a kind of Tamm/Shockley surface states in molecular structures \cite{Kout}) have been discussed by many authors in the context of zigzag graphene ribbons \cite{Fujita,Nakada,Peres1,Brey,prb2007}, where such states associate with decaying modes of the transverse electron motion. In distinction from the Tamm states, edge states of graphene are not fully localized: They decay in the (transverse) armchair direction. In the (longitudinal) zigzag direction, these states are delocalized and can be described by a superposition of propagating states.

By restricting ourselves to imaginary values of $\kappa^-_{j,0}=i\delta_j$, that is by the spectrum of decaying modes, we can rewrite the minus branch of Eq.~(\ref{8}) in the form 
\begin{equation}\label{12}
E^-=\pm \frac{\sinh(\delta_j/2)}{\sinh \delta_j(N+1/2)}, 
\end{equation}
where $\delta_j$ must satisfy
\begin{equation}\label{13}
\frac{\sinh\delta_j N}{\sinh \delta_j(N+1/2)}=2\cos(\xi_j/2),\quad j\ge j^*. 
\end{equation}

\begin{figure}
\includegraphics[width=0.4\textwidth]{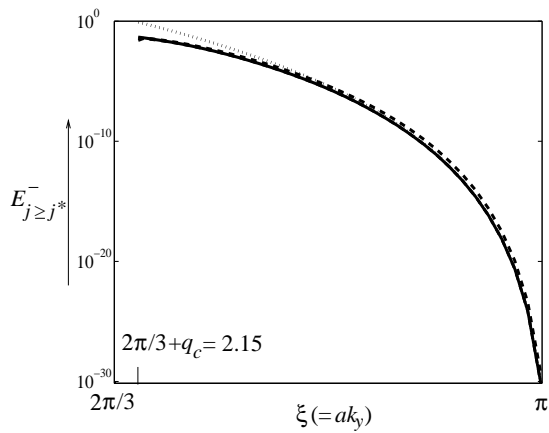}
\caption{ Energy of edge states, $E^-_{j\ge j^*}$, as a function of $\xi_j$. Exact dependence and Eq.~(\ref{15}) are represented by a single solid line; approximation (\ref{16}) for $\delta_j N >>1$ and $\pi/3-q_j<<1$ are shown by dashed and dotted lines, respectively.}
\end{figure}

The two equations above give the exact position of edge-state levels in the graphene spectrum. An approximate solution of Eqs.~(\ref{13}), $\delta_j =-2\ln\left[2\cos(\xi_j/2)\right]$,
yields the values of edge-state energies,
\begin{equation}\label{15}
E^-_{j\ge j^*}=\pm\frac{[2\cos(\xi_j/2)]^{-1}-2\cos(\xi_j/2) }{[2\cos(\xi_j/2)]^{-2(N+1)}-[2\cos(\xi_j/2)]^{2(N+1)}},
\end{equation}
which even in the logarithmic scale are indistinguishable from the exact solution; see Fig. 3.
For values of $q_j$, which are close to $\pi/3$ and under 
 the condition $\delta_j N >>1$, Eq.~(\ref{15}) simplifies to
\begin{eqnarray}\label{16}
E^-_{j\ge j^*}=&&
\delta_j\exp(-\delta_j N),\qquad \qquad \delta_j N >>1, \nonumber\\
&& \left[\pi- \pi j /({\cal N}+1)\right]^{2N},
 \pi/3-q_j<<1. 
\end{eqnarray}

The exponential behavior shown here has been observed previously in many numerical models, e.g., 
\cite{Nakada,Peres1,prb2007}, and an analytical description was given by Brey and Fertig \cite{Brey}. Qualitatively, the latter agrees with Eq.~(\ref{16}), but the functional form of the exponential factor is very different from our exact result. Further comments on this point can be found in Ref. \cite{Lyuba2}.

To summarize our findings, we have presented an exact quantitative description of a rectangular sheet of graphene which brings to light the crucial role of zigzag boundaries in determining the $\pi$ electron spectrum near the Fermi energy. This result devalues the concept of a "zero mode" with no dispersion. Secondarily, this solves a number of long standing problems, which have been the subject of a considerable computational and analytical effort. Not immediate but a straightforward application of the obtained solution is an accurate description of the spectrum of achiral graphene ribbons and carbon nanotubes near the point of neutrality \cite{Lyuba3}. Altogether this forms a new platform for arguable interpretation and modeling of the electronic properties of graphene and its daughter structures.

We are thankful to Linda Wylie and Stephen Macken for their valuable comments on the manuscript. The work was partly supported by Special Program of the Physics and Astronomy Section of National Academy of Science of Ukraine, Visby program of the Swedish Institute, and a grant from the Royal Academy of Science (KVA).

\end{document}